\documentclass[aps,pra,twocolumn,superscriptaddress]{revtex4-1}
\usepackage{amsmath,amsfonts,amssymb}
\usepackage{graphicx,float,calc}
\usepackage{color,bm}
\usepackage{ulem}
\usepackage[colorlinks,urlcolor=blue,citecolor=blue,linkcolor=blue]{hyperref}
\usepackage{url}

\begin{document}

\title{Simulating bosonic Chern insulators in one-dimensional optical superlattices}

\author{Yu-Lian Chen}
\affiliation{Guangdong Provincial Key Laboratory of Quantum Engineering and Quantum Materials,GPETR Center for Quantum Precision Measurement, Frontier Research Institute for Physics and SPTE, South China Normal University, Guangzhou 510006, China}
\author{Guo-Qing Zhang}
\affiliation{Guangdong Provincial Key Laboratory of Quantum Engineering and Quantum Materials,GPETR Center for Quantum Precision Measurement, Frontier Research Institute for Physics and SPTE, South China Normal University, Guangzhou 510006, China}
\author{Dan-Wei Zhang}
\email[]{danweizhang@m.scnu.edu.cn}
\affiliation{Guangdong Provincial Key Laboratory of Quantum Engineering and Quantum Materials,GPETR Center for Quantum Precision Measurement, Frontier Research Institute for Physics and SPTE, South China Normal University, Guangzhou 510006, China}
\author{Shi-Liang Zhu}
\email[]{slzhu@nju.edu.cn}
\affiliation{National Laboratory of Solid State Microstructures and School of Physics, Nanjing University, Nanjing 210093, China}
\affiliation{Guangdong Provincial Key Laboratory of Quantum Engineering and Quantum Materials,GPETR Center for Quantum Precision Measurement, Frontier Research Institute for Physics and SPTE, South China Normal University, Guangzhou 510006, China}

\date{\today}

\begin{abstract}
We study the topological properties of an extended Bose-Hubbard model with cyclically modulated hopping and on-site potential parameters, which can be realized with ultracold bosonic atoms in a one-dimensional optical superlattice. We show that the interacting bosonic chain at half filling and in the deep Mott insulating regime can simulate bosonic Chern insulators with a topological phase diagram similar to that of the Haldane model of noninteracting fermions. Furthermore, we explore the topological properties of the ground state by calculating the many-body Chern number, the quasiparticle energy spectrum with gapless edge modes, the topological pumping of the interacting bosons, and the topological phase transition from normal (trivial) to topological Mott insulators. We also present the global phase diagram of the many-body ground state, which contains a superfluid phase and two Mott insulating phases with trivial (a zero Chern number) and nontrivial topologies (a nonzero Chern number), respectively.

\end{abstract}

\maketitle

\section{Introduction}

Recently, the exploration of topological phases has become an important research area in condensed matter physics \cite{Hasan2010,XLQi2011,Bansil2016,Chiu2016,Armitage2018} and various artificial systems, such as topological photonic and mechanic systems \cite{LLu2014,Ozawa2019,Huber2016} and superconducting circuits \cite{Schroer2014,Roushan2014,XTan2018,XTan2019b}. In particular, ultracold atoms in optical lattices provide a powerful platform for quantum simulation of many-body physics and topological states of matter \cite{Jaksch1998,Greiner2002,LMDuan2003,Bloch2008,Lewenstein2007,DWZhang2018,Goldman2016,Cooper2019,DWZhang2011,YXu2019}. With the advances of synthetic gauge field and spin-orbit coupling for neutral atoms \cite{Dalibard2011,Goldman2014,Galitski2013,HZhai2015,Ruseckas2005,Osterloh2005,SLZhu2006,XJLiu2007}, various topological phases and phenomena have been achieved in cold atom experiments. For instance, the celebrated Su-Schrieffer-Heeger (SSH) model \cite{WPSu1979} has been realized in one-dimensional (1D) dimerized optical superlattices \cite{Atala2013}. The topological (geometric) pumping with ultracold fermionic or bosonic atoms in the optical superlattices has been demonstrated \cite{Thouless1983,YQian2011,LWang2013a,LWang2013b,FMei2014,DWZhang2015,Nakajima2016,Lohse2015,HILu2016,Schweizer2016}. The Harper-Hofstadter model \cite{Harper1955,Hofstadter1976} and Haldane model \cite{Haldane1988} for the (anomalous) quantum Hall effect have been realized in two-dimensional (2D) optical lattices \cite{Miyake2013,Aidelsburger2013,Aidelsburger2014,Jotzu2014,LBShao2008,CWu2008a}, where the topological Chern numbers characterizing the band topology were measured from the dynamical response of atomic gases \cite{Aidelsburger2014}. The 2D topological bands were also realized in optical Raman lattices \cite{XJLiu2014,ZWu2016}. The chiral edge states, as one hallmark of Chern insulators, were observed with cold atoms in synthetic Hall ribbons \cite{Mancini2015,Stuhl2015}.

Although these works on studies of topological phases with ultracold atoms are at the single-particle level, several recent experiments have been done to explore topological phenomena in the interacting regime \cite{Schweizer2016,Tai2017,Leseleuc2019}. In addition, it has been theoretically revealed that the combination of band topology and many-body interactions (correlation effects) can give rise to exotic interacting topological phases \cite{Cooper2008,Goldman2016,Rachel2018,Daniel2018,Daniel2019,Magnifico2019a,Magnifico2019b}, which include factional and integer quantum Hall effects of bosons \cite{Sorensen2005,Senthil2013,Furukawa2013,YCHe2015,Sterdyniak2015,TSZeng2016,WLiu2019} and the topological Mott insulator \cite{Raghu2008,YZhang2009,Yoshida2014,Herbut2014,Amaricci2016}. The bosonic integer quantum Hall state is a $U(1)$ symmetry-protected topological phase \cite{XChen2012,ZXLiu2014}, which could be realized with interacting bosonic atoms in 2D optical lattices under an artificial magnetic field as extended Bose-hubbard models \cite{YCHe2015,Sterdyniak2015,TSZeng2016}. The topological Mott insulator was first predicted in a 2D honeycomb lattice of interacting fermions \cite{Raghu2008}, and then was shown to be a generic class of interaction-induced topological insulators for interacting fermions or bosons in different dimensions \cite{YZhang2009,Yoshida2014,Herbut2014,Amaricci2016,Dauphin2012,Dauphin2016,Amaricci2015,Imriska2016,Barbarino2019,Irsigler2019,SLZhu2013,XDeng2014,Kuno2017,Grusdt2013,ZXu2013,TLi2015,HHu2017,Stenzel2019}. Several schemes have been proposed to realize the topological Mott insulating phase by using interacting bosonic or fermionic atoms in 2D optical lattices \cite{Dauphin2012,Dauphin2016} and 1D optical superlattices \cite{SLZhu2013,XDeng2014,Kuno2017,ZXu2013,Grusdt2013,TLi2015,HHu2017,Stenzel2019}. Notably, it has been shown that a topological phase characterized by a $Z_2$ topological invariant (quantized Berry's phase) for strongly repulsive bosons in the half-filled Mott insulator can be induced by replacing free fermions with strongly interacting bosons in the SSH model \cite{Grusdt2013}. An interesting but less explored question is how to simulate the topological physics of 2D bosonic Chern insulators in a simple Bose-Hubbard type of model realizable in 1D optical lattices \cite{SLZhu2013,XDeng2014,Kuno2017}.

In this paper, we study the topological properties of an extended Bose-Hubbard model with cyclically modulated hopping and on-site potential parameters in 1D dimerized optical superlattices, based on numerical exact diagonalization (ED) \cite{JMZhang2010,Weinberg2017} and density matrix renormalization group (DMRG) methods \cite{White1992,Schollwoeck2011}. Our main results are as follows: (i) We show that this 1D interacting bosonic system at half filling and in the Mott insulating regime can simulate bosonic Chern insulators with a topological phase diagram similar to that of the celebrated Haldane model. For hardcore bosons that can be mapped to free fermions, this system corresponds to the generalized SSH model of noninteracting fermions studied in Ref. \cite{LLi2014}. We show that the interacting topological phases can preserve for soft bosons (in the absence of the mapping) with the many-body Mott gap. (ii) We explore the topological properties of the ground state by calculating the many-body Chern number, the quasiparticle energy spectrum with gapless edge modes, the topological pumping of the interacting bosons, and the topological phase transition from a trivial Mott insulator to a topological Mott insulator. We also show the breakdown of topological pumping of edge modes for weakly interacting bosons in the Mott insulating phase. (iii) Finally, we obtain a global ground-state phase diagram consisting of a superfluid phase and two Mott insulating phases.

The rest of this paper is organized as follows. In Sec. \ref{sec2}, we propose the extended 1D Bose-Hubbard model in a dimerized optical superlattice. In Sec. \ref{sec3}, we investigate the topological properties of the many-body ground state at the half filling in the model and show that this system can simulate an analog of Chern insulators of strongly interacting bosons. Finally, a brief summary is presented in Sec. \ref{sec4}.

\section{Model}\label{sec2}

Let us begin by considering a 1D optical superlattice consisting of single-component interacting bosonic atoms with two sites (sublattices A and B) per unit cell \cite{Atala2013,Nakajima2016,Lohse2015,HILu2016,Schweizer2016}. In the tight-binding region, the system can be described by the following extended Bose-Hubbard Hamiltonian
\begin{equation}\label{eq-ham}
\begin{split}
&\hat{H}=\hat{H}_{0}+\frac{U}{2}\sum_j\hat{n}_j(\hat{n}_j-1), \\
&\hat{H}_{0}=\sum_{j\in \text{odd}}(t_1\hat{a}_j^{\dagger}\hat{a}_{j+1}+h.c.)+\sum_{j\in \text{even}}(t_2\hat{a}_j^{\dagger}\hat{a}_{j+1}+h.c.)\\
&~~~~~~+\sum_{j\in \text{odd}}\mu_A\hat{n}_{j}+\sum_{j\in \text{even}}\mu_B\hat{n}_{j},
\end{split}
\end{equation}
where $U$ denotes the on-site atomic interaction energy, $\hat{n}_{j}=\hat{a}_j^{\dagger}\hat{a}_j$ is the number operator in the $j$-th lattice, with $\hat{a}_{j}^{\dagger}$ and $\hat{a}_{j}$ corresponding the bosonic creation and annihilation operators, respectively. Here the single-particle Hamiltonian $H_0$ describes a generalized SSH chain with the intra- and inter-cell hopping energies
\begin{equation}\label{eq-hopping}
t_1=t(1+\delta \cos\beta), ~~t_2=t(1-\delta\cos\beta),
\end{equation}
where the characteristic hopping strength $t=1$ is set as the energy unit hereafter, $\delta$ represents the dimerization strength and $\beta$ is a cyclical parameter varying from $-\pi$ to $\pi$ continuously, and $\mu_A$ and $\mu_B$ are the alternating on-site potentials on the sublattices A and B. In addition, the alternating on-site potentials $\mu_A$ and $\mu_B$ in $H_0$ are parameterized as
\begin{equation}\label{eq-onsite}
\mu_A=g_A+h\cos(\beta+\phi), ~~\mu_B=g_B+h\cos(\beta-\phi),
\end{equation}
where $g_A$, $g_B$, $h$, and $\phi$ are parameters tunable by external laser fields \cite{Atala2013,Nakajima2016,Lohse2015,HILu2016,Schweizer2016}. Here $\phi\in[-\pi,\pi]$ is closely linked to whether we get symmetric ($\phi=0$) or antisymmetric ($\phi=\pm\pi/2$) alternating on-site potentials in the case of $g_A=g_B=0$.  

In the noninteracting case \cite{LLi2014}, $\hat{H}_{0}$ can be rewritten as $\hat{H}_{0}=\sum_k\psi_k^{\dagger}h_0(k)\psi_k$ in momentum space with the Bloch Hamiltonian
\begin{equation}\label{eq-h0}
h_0(k)=\begin{bmatrix}

      \mu_A  & t_1+t_2e^{-ik}\\

      t_1+t_2e^{ik} & \mu_B
\end{bmatrix}
,
\end{equation}
where $\psi_k^{\dagger}=(\hat{a}_k^{\dagger},\hat{b}_k^{\dagger})$. Diagonalizing the Hamiltonian and defining $g_+=(g_A+g_B)/2$ and $g_-=(g_A-g_B)/2$, we get two Bloch bands
\begin{equation}
E_{\pm}(k)=g_++h\cos{\beta}\cos{\phi}\pm\sqrt{\Delta},
\end{equation}
where $\Delta=(g_-+h\sin{\beta}\cos{\phi})^2+2(1+\cos{k})+2(1-\cos{k})\delta^2\cos^2{\beta}$. Hereafter we drop the irrelevant energy constant $g_+$ by setting $g_A=-g_B=g$, such that $g_-=g$.

It is well-known that two distinct phases exhibit in the original SSH model when $|t_1|<|t_2|$ and $|t_1|>|t_2|$, respectively, with a topological transition at $|t_1|=|t_2|$ \cite{WPSu1979,DWZhang2018}. The generalized SSH model described by the Hamiltonian $h_0(k)$ in Eq. (\ref{eq-h0}) can be mapped to the 2D Haldane model in the momentum space \cite{LLi2014}, where the topological Chern number is defined in an extended 2D parameter space spanned by the momentum $k$ and the additional modulation phase $\beta$. The mapping is achieved by identifying $k$ with $k_x$, $\beta$ with $k_y$, $\phi$ with $\varphi$, and $g/h$ with $M_z$, where $k_{x,y}$, $\varphi$, and $M_z$ denote the two quasimomenta, the next nearest-neighbor hopping phase, and the effective Zeeman filed in the Haldane model \cite{Haldane1988,LLi2014}.

The topological phase diagram of the 1D single-particle Hamiltonian for free fermions ($\hat{H}_0$ with modulated hopping parameters for simulating the Haldane model) have been studied in Ref. \cite{LLi2014}. In the following section, we explore the topological and quantum phases of the interacting bosons in this generalized SSH chain, and we show that this 1D system in the Mott insulating phase can simulate a bosonic analog of fermionic Chern insulators in the 2D Haldane model. Note that the quantum phases in a similar 1D Bose-Hubbard model (without the cyclically modulated hopping and on-site potential parameters) was studied in Ref. \cite{Rousseau2006}. Here we focus on the topological properties of the Mott insulating ground-state by the ED and DMRG simulations, which are partially performed using the open-source Quspin library \cite{Weinberg2017} and ITensor library \cite{ITensor}, respectively. The DMRG is regarded as the one of the best methods for the simulation of 1D strongly correlated quantum lattice systems \cite{White1992,Schollwoeck2011}. In the numerical DMRG calculations, we truncate the local bosonic basis up to a maximal occupation number (the maximum number of bosons per site) and increase the maximal occupation during our calculation until reaching convergence.

\section{Topological properties}\label{sec3}

In this section, we systematically investigate the topological properties of the 1D generalized Bose-Hubbard model in Eq. (\ref{eq-ham}). In particular, we focus on the ground state at half filling, in which case the particle number is $N_a=L/2$ with the lattice length $L$. We first consider the excitation gap and the many-body Chern number to characterize the insulating ground state of the system. We then study the quasiparticle energy spectrum with gapless edge modes and the topological pumping of the interacting bosons in the deep Mott insulator regime. Finally, we reveal the topological phase transition from a trivial Mott insulator to a topological Mott insulator and present the global phase diagram of the ground state.

\subsection{Excitation gap and Chern number}
We first perform ED calculations of the model Hamiltonian (\ref{eq-ham}) for small-size lattices under periodic boundary conditions (PBCs) to obtain the eigenenergies and the Chern number of the many-body ground state. At half filling and under PBCs, the ground state of the system is non-degenerate and separated from the first excited state by an excitation gap
\begin{equation}
\Delta_{\text{ex}}=E_{e}-E_g,
\end{equation}
where $E_g$ and $E_e$ denote the energies of the ground and first excited states for fixed $N_a=L/2$, respectively.

\begin{figure}[htbp]
\centering
\includegraphics[width=6.5cm]{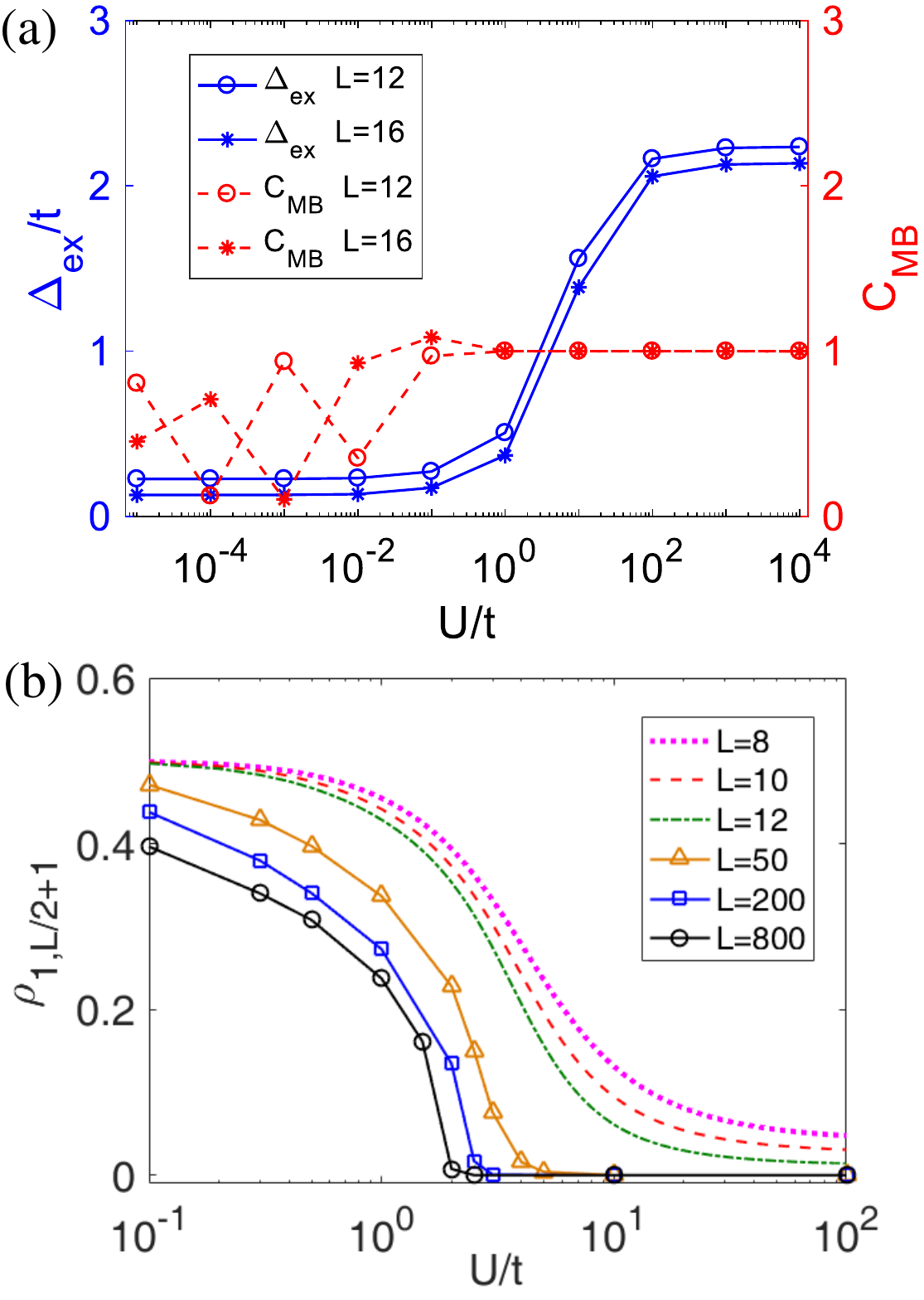}
\caption{(Color online). (a) The excitation gap $\Delta_{\text{ex}}$ and the many-body Chern number $C_{\text{MB}}$ as a function of the interaction strength $U$. The lattice sites $L=2N_a=12$ and $16$ are used in the ED of the model Hamiltonian (\ref{eq-ham}) under PBCs. (b) The single-particle density matrix $\rho_{1,L/2+1}$ as a function of $U$ for $L=8,10,12$ from the ED and $L=50,200,800$ from the DMRG. Other parameters are $\beta=0$, $g/h=0$, and $\phi=-\pi/2$, $h=1$, and $\delta=-0.5$.}
\label{fig:1}
\end{figure}

\begin{figure}[htbp]
\centering
\includegraphics[width=6.5cm]{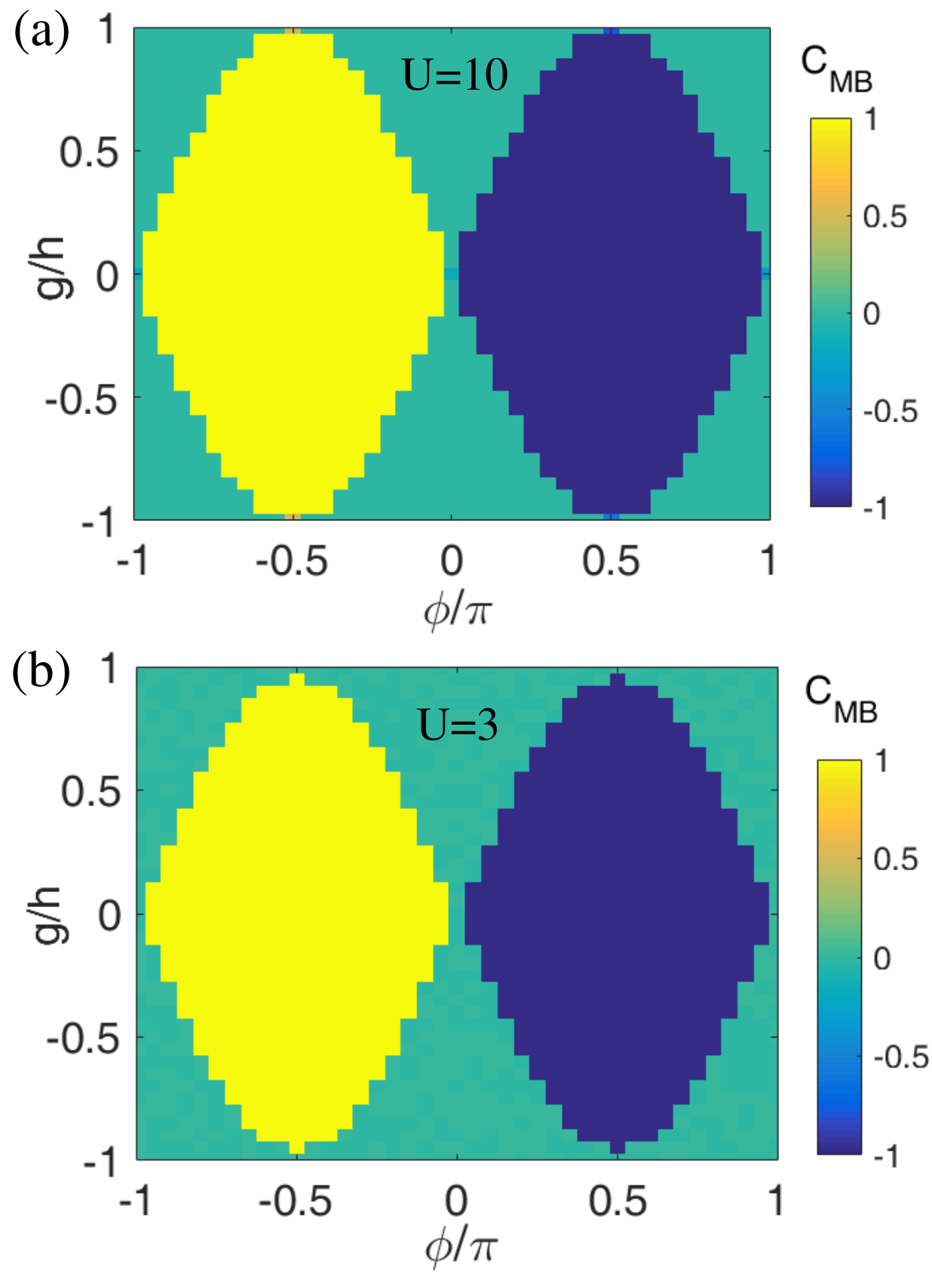}
\caption{(Color online). Topological phase diagrams obtained from $C_{\text{MB}}$ with respect to $g/h$ and $\phi$ for $U=10$ (a) and $U=3$ (b) at half filling. The phase diagrams simulate the topological phase diagram of the Haldane model of free fermions, with $C_{\text{MB}}=0$ and $1$ denoting the trivial and topological insulating phases, respectively. Other parameters are $h=1$, $\phi=-\pi/2$, $\delta=-0.5$, and $L=12$.}
\label{fig:2}
\end{figure}

The calculated excitation gap $\Delta_{\text{ex}}$ as a function of the interaction strength $U$ for $L=12$ and $16$ is shown in Fig. \ref{fig:1}(a). It is clear that $\Delta_{\text{ex}}$ is virtually flat for small $U$ and then increases monotonically with respect to $U$ and finally saturates at a finite value when $U$ is large enough. The ED results imply that if $L$ is large enough, the excitation gap decreases to zero as $U$ drops below a critical value, namely $U_{c}$. The cold bosonic atomic system is then in the superfluid phase \cite{Jaksch1998,Greiner2002,Rousseau2006}. The superfluid state can be characterized by a finite value of the single-particle density matrix under PBCs \cite{JMZhang2010}, such as $\rho_{1,L/2+1}=\langle \Psi_{g} |a_1^{\dagger} a_{L/2+1}|\Psi_{g}\rangle$ in our case, where $|\Psi_{g}\rangle$ denotes the ground-state wave function. The single-particle density matrix associated with the many-particle ground state captures the momentum distribution of the interacting bosons. The presence of a condensate in the superfluid phase at zero temperature is associated with an off-diagonal (quasi-)long-range order indicated by the results that $\rho_{1,L/2+1}$ remains finite at large $L$. The typical results of $\rho_{1,L/2+1}$ in Fig. \ref{fig:1}(b) show that the superfluid (Mott insulator) phase forms for small (large) $U$, which exhibits the dependence of the lattice size $L$ and takes the critical value $U_{c}\approx2$ when $L$ is up to 800.

When $U\rightarrow\infty$, the atoms become hardcore bosons and the Bose-Hubbard model can be mapped to the free fermion model with the two-band Bloch Hamiltonian $h_0$ in Eq. (\ref{eq-h0}), where the band gap is $2t=2$ for the chosen parameters. The small deviations of $\Delta_{\text{ex}}$ between the ED results and the expected values in the large and small $U$ limits ($\Delta_{\text{ex}}=0$ for $U<U_{c}$ and $\Delta_{\text{ex}}=2$ for large $U$) are due to the finite-size effect and can be  decreased by enlarging the lattice size.

We calculate the Chern number to characterize topology of the many-body ground state with finite excitation gaps. For free fermions in Bloch bands, the Chern number is usually defined as an integration over the occupied states in momentum space \cite{Thouless1982}. However, the definition is no longer applicable for bosons as many bosons can occupy the same momentum state. For the interacting bosonic system, we suppose the ground state $|\Psi_g(j, \theta, \beta)\rangle$ is separated from the excited states and depends on the parameters $\theta$ and $\beta$ through the twisted PBC \cite{QNiu1985}: $|\Psi_g(j+L, \theta, \beta)\rangle=e^{i\theta}|\Psi_g(j, \theta, \beta)\rangle$, where $j=1,\cdots,L$, and $\theta\in[-\pi,\pi]$ is an imposed twisted phase. The many-body Chern number is given by \cite{QNiu1985}
\begin{equation}\label{ChN}
C_{\text{MB}}=\frac{1}{2\pi}\int_{-\pi}^{\pi} d\theta \int_{-\pi}^{\pi} d\beta(\partial_\beta A_\theta-\partial_\theta A_\beta),
\end{equation}
where $A_\mu=i\langle{\Psi_g(\theta,\beta)|\partial_\mu|\Psi_g(\theta,\beta)}\rangle$ $(\mu=\beta,\theta)$ is the Berry connection. Note that the Chern number defined in Eq. (\ref{ChN}) is not related to a quantized Hall conductivity, but can (could not) be related to the topological pumping for strongly (weakly) interacting bosons in 1D optical lattices [see Eq. (\ref{COM})].

We numerically calculate $C_{\text{MB}}$ by evaluating the Berry curvature $F(\theta,\beta)=\partial_\beta A_\theta-\partial_\theta A_\beta$ for a discretized manifold spanned by $\theta$ and $\beta$ \cite{Fukui2005}. Notably, we define the bulk gapped phase under the twisted PBC if the many-body ground state remains gapped in the whole $\theta$-$\beta$ parameter space. We depict $C_{\text{MB}}$ as a function of $U$ at half filling in Fig. \ref{fig:1}(a). We find that $C_{\text{MB}}$ is quantized to unit for the Mott insulator states when $U>U_{c}$, while $C_{\text{MB}}$ is unquantized (and actually is not well-defined) due to the gapless for the superfluid phase.

We emphasize in the topological properties of the Mott insulator states and obtain a topological phase diagram with respect to $\phi$ and $g/h$ from calculated $C_{\text{MB}}$ for $U=10$, as shown in Fig. \ref{fig:2}(a). There are a trivial Mott insulating phase with $C_{\text{MB}}=0$ and two topological Mott insulating phases with $C_{\text{MB}}=\pm1$. The topological phase transition will be addressed later. It is clear that the phase diagram obtained for the strongly interacting bosons in the 1D dimerized lattice is similar to the free-fermion case in the 2D Haldane model. Thus, we can use ultracold bosons in the 1D optical superlattice to simulate the bosonic Chern insulators, which can be viewed as anomalous quantum Hall effect of interacting bosons in 2D lattices. Notably, this topological phase is not a correlated 2D phase of bosons and can be understood for 1D hardcore bosons via mapping them to free fermions by virtue of the Jordan-Wigner mapping. Since the Chern number in this system is protected by the many-body Mott gap, the topological phase diagram remains the same even for smaller interactions, such as the result for $U=3$ shown in Fig. \ref{fig:2}(b). In the following, we further study the topological and quantum phases of the model system. Without loss of generality, we fix the parameters $h=1$, $\delta=-0.5$, and $\phi=-\pi/2$. We first study the case of $g/h=0$ and then obtain the global phase diagram with various $g/h$.

\begin{figure}[htbp]
\centering
\includegraphics[width=8.5cm]{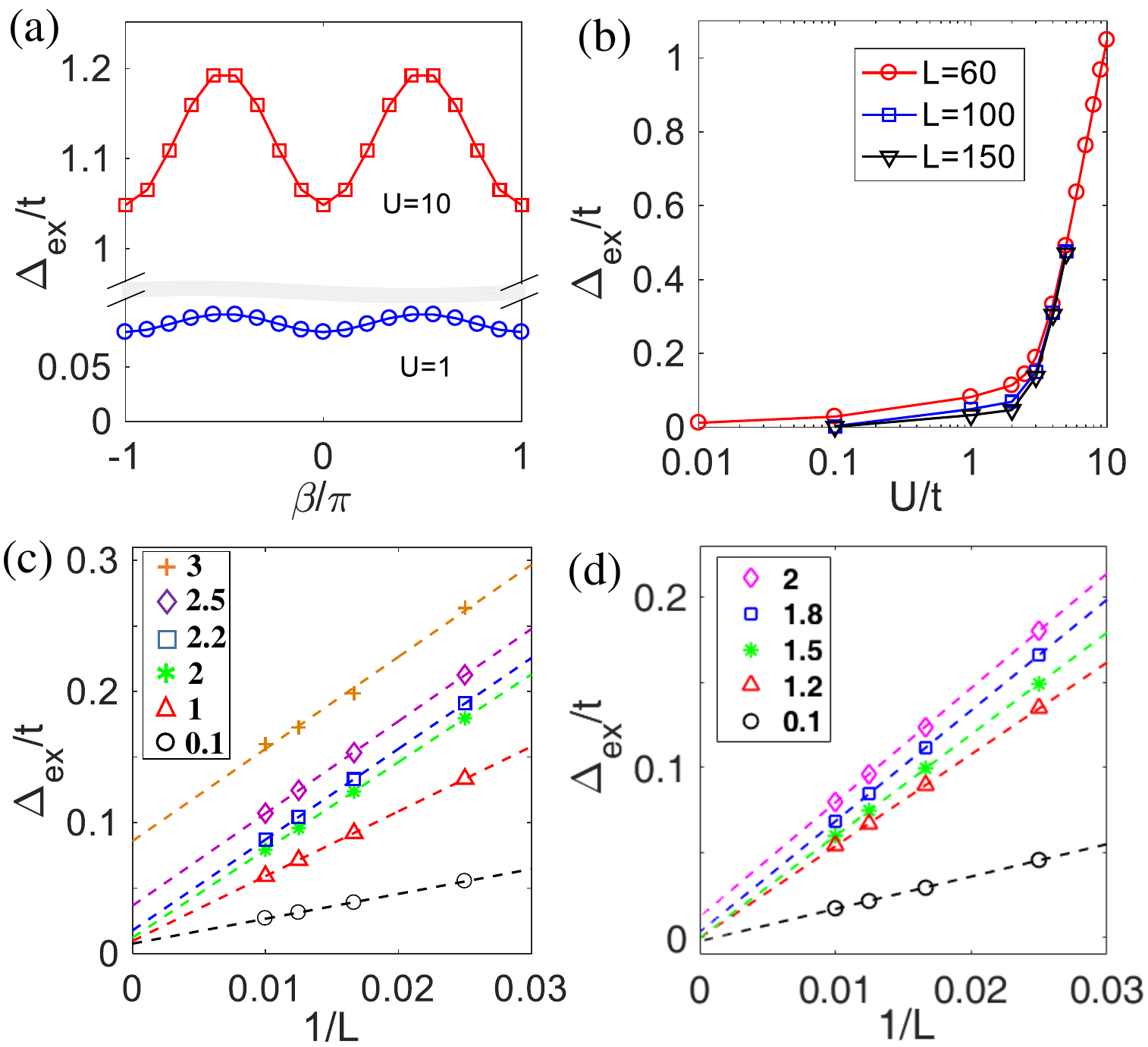}
\caption{(Color online). (a) The excitation gap $\Delta_{\text{ex}}$ as a function of the phase parameter $\beta$ for the lattice length $L=60$, $g/h=0$, and $U=1,10$. (b) $\Delta_{\text{ex}}$ as a function of $U$ for $\beta=g/h=0$ and $L=60,100,150$. (c,d) The finite-size scaling for various $U$ (the labels) with fixed $g/h=0$ and $g/h=0.5$, respectively. Other parameters are $h=1$, $\delta=-0.5$, and $\phi=-\pi/2$.}
\label{fig:3}
\end{figure}

Before proceeding, we further confirm the critical interaction strength $U_{c}\approx2$ for the quantum phase transition from the superfluid to Mott insulating phases, based on the DMRG calculation of the excitation gap $\Delta_{\text{ex}}$. In Fig. \ref{fig:3}(a), we show $\Delta_{\text{ex}}$ as a function of the phase parameter $\beta$ for $U=1,10$ and find that $\Delta_{\text{ex}}$ takes the minimum value when $\beta=0,\pm\pi$ for $g/h=0$. Thus, we consider the excitation gap with fixed $\beta=0$ and calculate $\Delta_{\text{ex}}$ as a function of $U$. As shown in Fig. \ref{fig:3}(b), with increasing lattice size $L=60,100,150$ for $\Delta_{\text{ex}}(U)$, we find a critical interaction strength approaching $U\approx2$ for the same parameters in Fig. \ref{fig:1}(a). To further determine $U_{c}$ in the thermodynamic limit, we perform the finite-size scaling of $\Delta_{\text{ex}}$ for various $U$, which should approach to nonzero (zero) in the Mott insulating (superfluid) phase. The finite-size scaling in Fig. \ref{fig:3}(c) shows that $\Delta_{\text{ex}}$ in the $L\rightarrow \infty$ limit tends to be finite when $U\gtrsim2$, which gives the critical interaction strength $U_{c}\approx2$ for the parameters in Fig. \ref{fig:1}(a). By the same procedure, we can determine $U_{c}$ for the cases of $g/h\neq0$. For instance, the finite-size scaling analysis in Fig. \ref{fig:3}(d) for $g/h=0.5$ gives $U_c\approx1.8$. Notably, for the 1D Bose-Hubbard model, interactions can be reduced a lot and the system still maintains a Mott insulating phase \cite{Kuhner2000}.

\begin{figure}[htbp]
\centering
\includegraphics[width=8.5cm]{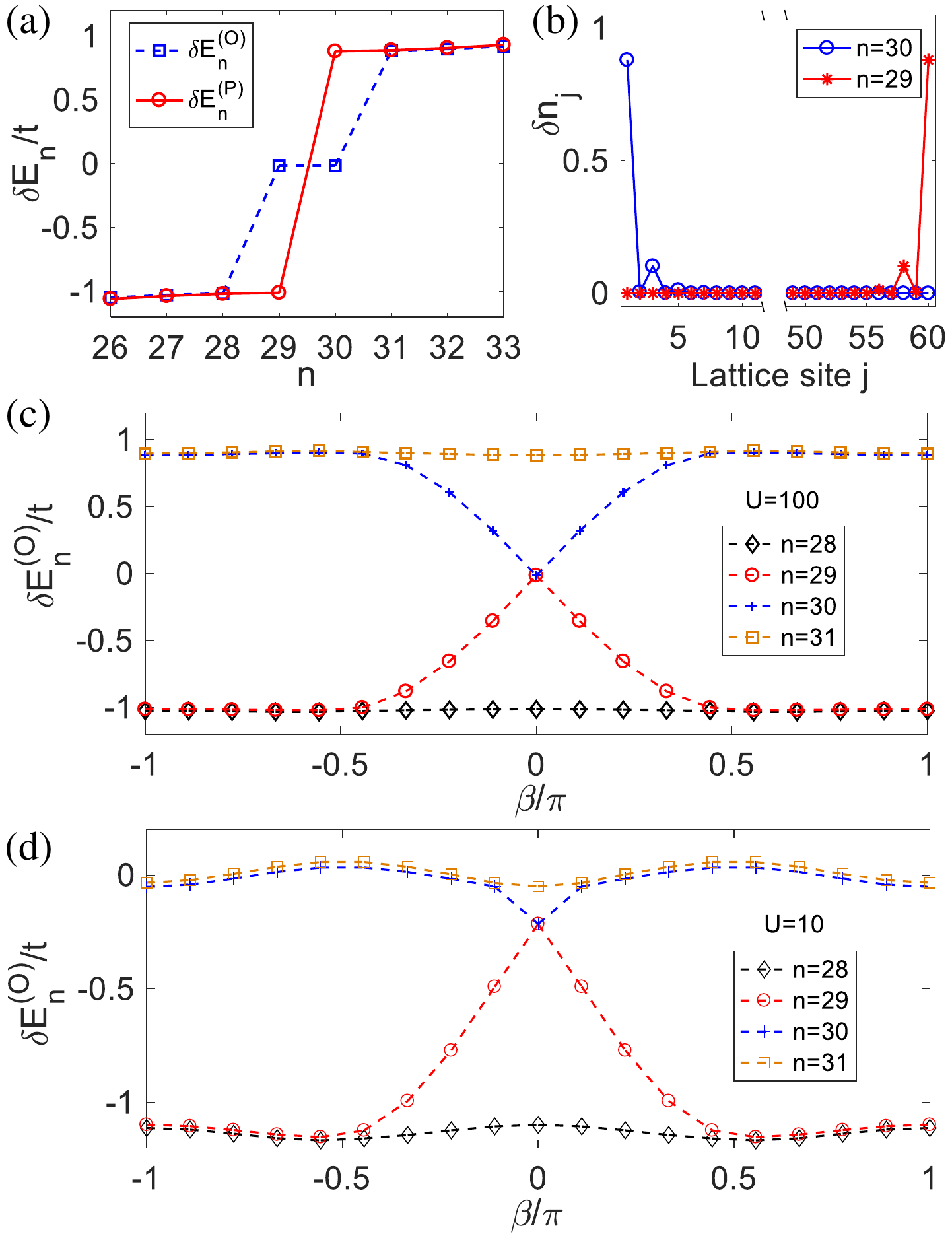}
\caption{(Color online). (a) The dependence of the quasiparticle energy spectrum $\delta E_n$ on the particle number $N_a=n$ under PBCs [$\delta E_n^{(P)}$] or OBCs [$\delta E_n^{(O)}$]. (b) The density distribution of the two in-gap edge modes. (c,d) The edges of the lower [$\delta E_{28}^{(O)}$] and the upper [$\delta E_{31}^{(O)}$] branches of the energy spectrum and the two in-gap modes [$\delta E_{29}^{(O)}$ and $\delta E_{30}^{(O)}$] as functions of the phase $\beta$ under OBCs. The other parameters are $L=60$, $h=1$, $\beta=0$, $g/h=0$, $\delta=-0.5$, $\phi=-\pi/2$, in (a-d), and $U=100$ in (a-c) and $U=10$ in (d).}
\label{fig:4}
\end{figure}

\subsection{In-gap edge states}

The appearance of in-gap edge states at the boundary is a hallmark of topologically nontrivial systems. To further study the properties of the bosonic topological Mott insulators in our model, we calculate the quasiparticle energy spectrum  \cite{SLZhu2013}
\begin{equation} \label{En}
\delta E_n^{(O,P)}=E_{n+1}^{(O,P)}-E_n^{(O,P)},
\end{equation}
where $E_n^{(S)}$ is the ground-state energy of the system with $N_a=n$ bosons, with $S=O,P$ denoting the open boundary conditions (OBCs) and PBCs, respectively. Here $\delta E_n^{(S)}$ represents the extra energy needed to add an atom to the system with $n$ atoms. The quasiparticle energy spectra for the lattice length $L=60$ and strong interaction strength $U=100$ under OBCs and PBCs are shown in Fig. \ref{fig:4}(a). It is clear that the spectrum splits into two branches separated by a finite gap near the half filling with two degenerate in-gap states under OBCs. The distribution of a quasiparticle in the presence of $n$ bosons filling in the lattice can be defined as \cite{HGuo2011}
\begin{equation}\label{nj}
\delta n_j=\langle \psi_{n+1}^{g}|\hat{n}_j|\psi_{n+1}^g\rangle-\langle \psi_{n}^{g}|\hat{n}_j|\psi_{n}^g\rangle,
\end{equation}
where $|\psi_n^g\rangle$ denotes the ground-state wave function of the system with $n$ bosons. We numerically calculate the density distribution of the two in-gap modes and find that they are localized near the two edges of the 1D lattice, as shown in Fig. \ref{fig:4}(b). This indicates that the ground state and first excited state differ by just edge excitations in the topological phase of the system under OBCs.

Furthermore, we show the energy spectra of the quasiparticles as functions of the phase $\beta$ under OBCs in Figs. \ref{fig:4}(c,d). There are two branches of edge modes (that are degenerate in energy at $\beta=0$) inside the bulk gap, which connect the lower and the upper bulk spectra when $\beta$ continuously varies from $-\pi$ to $\pi$. In the hardcore boson limit, the energy spectrum of the quasiparticles exhibits the particle-hole symmetry and has a bulk gap of $2$, which is similar to the free-fermion case. For the strong interacting case of $U=100$ shown in Fig. \ref{fig:4}(c), one can find that the energy spectrum slightly deviates the particle-hole symmetry with a bulk gap about $1.8$. In general, the particle-hole symmetry in the quasiparticle energy spectrum does not hold for soft bosons, which can be seen clearly from Fig. \ref{fig:4}(d) for the case of $U=10$. However, the topological properties can exhibit for soft-core bosons.

\subsection{Topological pumping of interacting bosons}

We now consider the topological pumping of the interacting bosons in the lattice. Topological charge pumping enables a quantized motion of particles through an adiabatic cyclic evolution in parameter space of the Hamiltonian \cite{Thouless1983}. The exact quantization of the transported particles in the thermodynamic limit is purely determined by the topology of the pump cycle. Especially, the pumped particles in an insulating state can be concerned with an integral of the Berry curvature over a closed surface that is quantized to be the Chern number. Experimentally, the topological (geometry) pumping has been demonstrated with cold atoms in 1D dimerized optical lattices from measuring the drift of the center-of-mass of atomic gases \cite{Nakajima2016,Lohse2015,HILu2016,Schweizer2016}.

\begin{figure}[htbp]
\centering
\includegraphics[width=8cm]{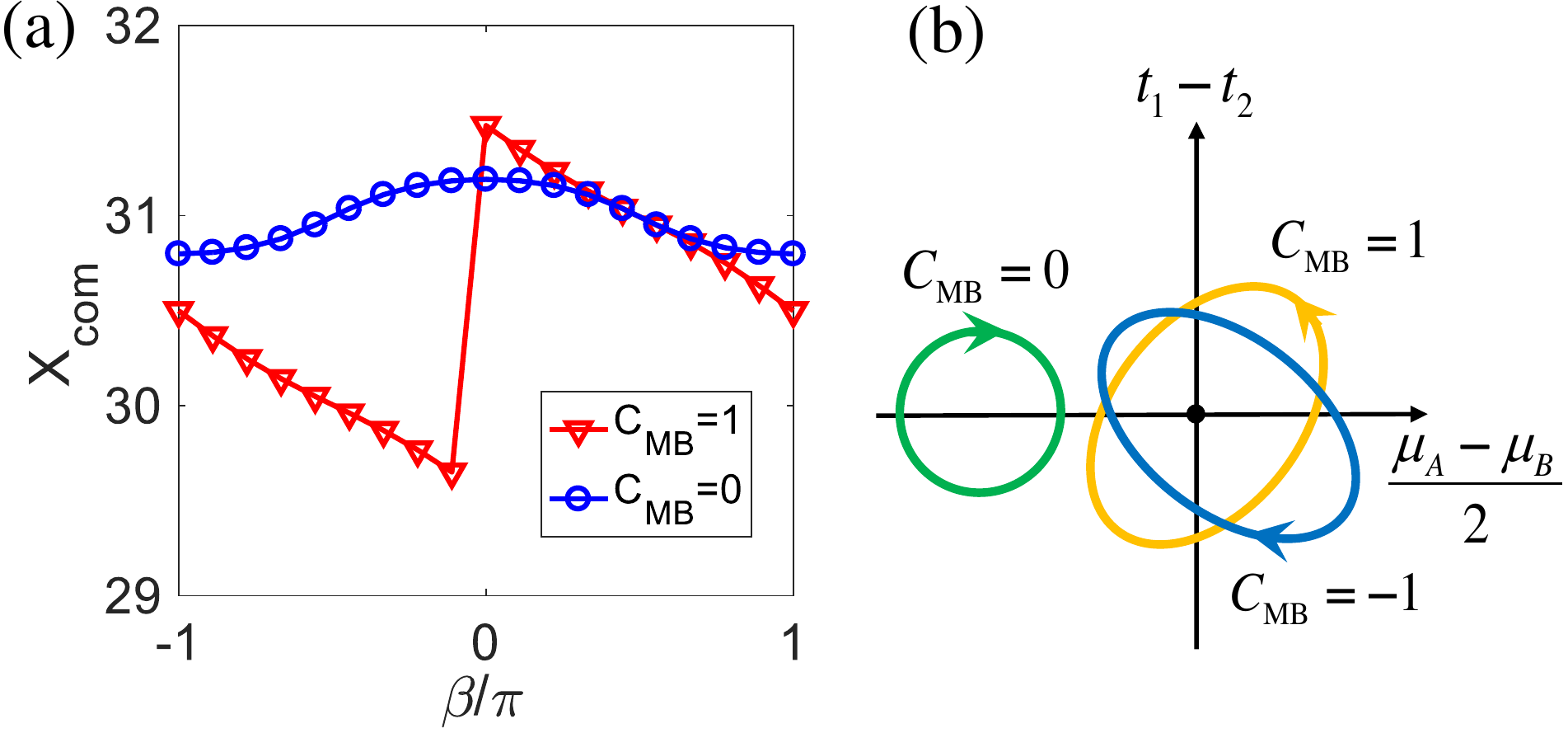}
\caption{(Color online). (a) The center-of-mass $X_{\text{COM}}$ as a function of the pump parameter $\beta$ for two examples of $C_{\text{MB}}=1$ ($g/h=0$ and $\phi=-\pi/2$) and $C_{\text{MB}}=0$ ($g/h=1$ and $\phi=0$) for $U=10$. (b) Schematic of three typical pump cycles for $C_{\text{MB}}=0,\pm1$ in the limit of hardcore bosons.}
\label{fig:5}
\end{figure}

For OBCs, we can calculate the center-of-mass of the ground state with respect to the adiabatic parameter $\beta$, which is given by
\begin{equation}
X_{\text{COM}}(\beta)=\frac{\sum_j j\langle{\Psi_g(\beta)|\hat{n}_j|\Psi_g(\beta)}\rangle}{\sum_j \langle{\Psi_g(\beta)|\hat{n}_j|\Psi_g(\beta)}\rangle}.
\end{equation}
Similar as that for noninteracting fermions \cite{LWang2013a,LWang2013b}, for the strongly interacting bosons during a pump cycle, the shift of the center-of-mass $\delta X_{\text{COM}}$ related to the Chern number is given by \cite{Schweizer2016}
\begin{equation}\label{COM}
\delta X_{\text{COM}}=\int_{-\pi}^{\pi}\frac{\partial X_{\text{COM}}(\beta)}{\partial\beta}d\beta.
\end{equation}
One has the relation $\delta X_{\text{COM}}=C_{\text{MB}}/d$ in the hardcore and $L\rightarrow\infty$ limit \cite{Hayward2018,Stenzel2019}, where $d=2a$ with the lattice spacing $a\equiv1$. In Fig. \ref{fig:5}(a), we show $X_{\text{COM}}$ as a function of $\beta$ for the lattice length $L=60$, with two examples of $C_{\text{MB}}=1$ and $C_{\text{MB}}=0$ in Fig. \ref{fig:2}, respectively. When $C_{\text{MB}}=1$ ($C_{\text{MB}}=0$), $X_{\text{COM}}$ exhibits a (no) jump of nearly one unit cell when $\beta$ varies from $-\pi$ to $\pi$, which gives rise to $\delta X_{\text{COM}}/d\approx1=C_{\text{MB}}$ ($\delta X_{\text{COM}}/d=C_{\text{MB}}=0$) as expected. The jump in $X_{\text{COM}}(\beta)$ corresponds to the shift of an occupied edge mode from one side of the lattice to the other under OBCs, which is the bulk-edge correspondence in topological pumping \cite{Hatsugai2016}.

\begin{figure}[htbp]
\centering
\includegraphics[width=8.5cm]{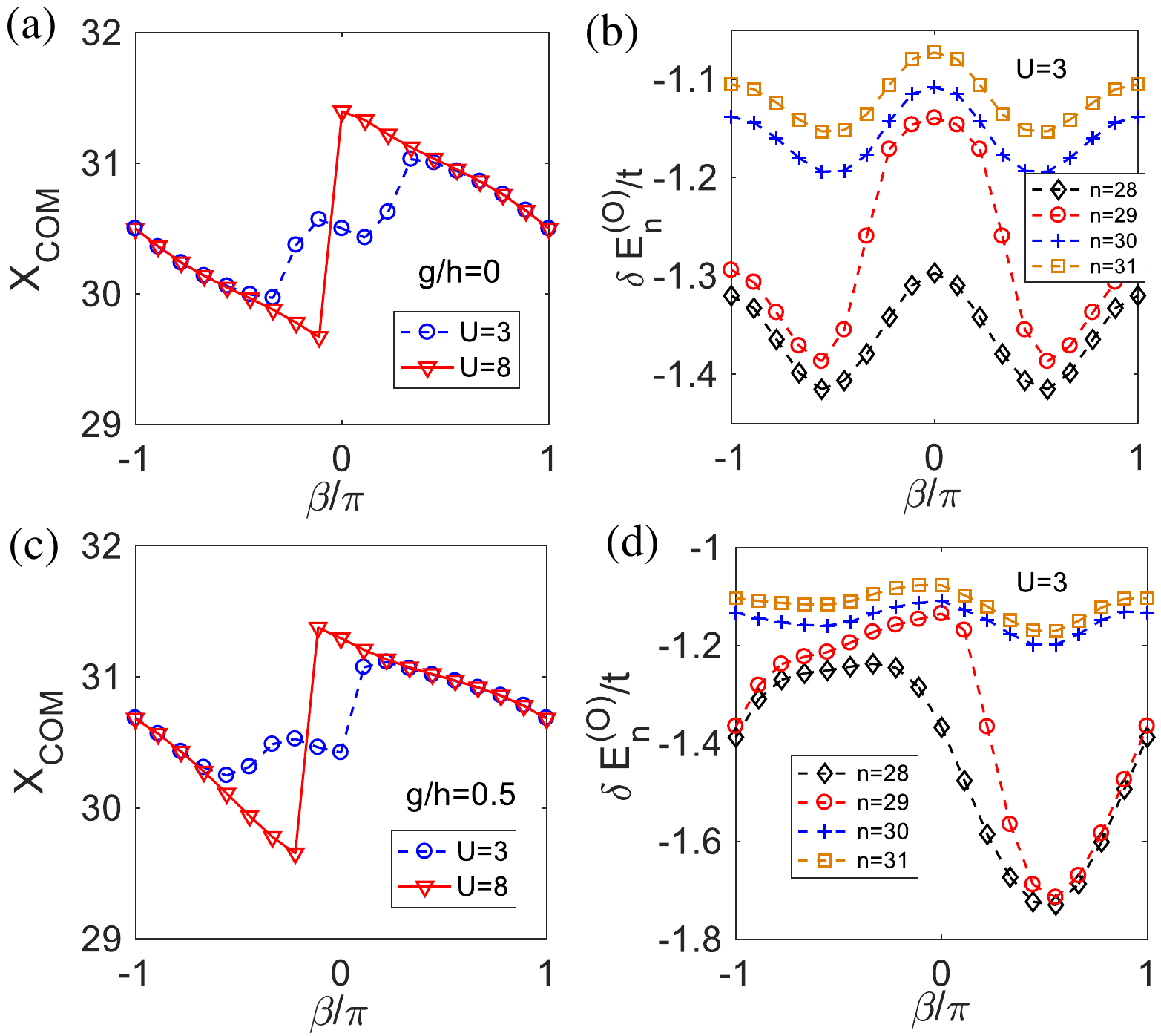}
\caption{(Color online). (a,c) The center-of-mass $X_{\text{COM}}$ as a function of the pump parameter $\beta$ with $U=3,8$ for $g/h=0$ and $g/h=0.5$, respectively. (b,d) The corresponding energy spectra for $U=3$ under OBCs. Other parameters are $\phi=-\pi/2$ and $L=60$.}
\label{fig:6}
\end{figure}

In the hardcore boson limit, the topological pumping can be simply understood from a single-particle picture. The hardcore bosons can be mapped to free spinless fermions via the Jordan-Wigner transformation \cite{Cazalilla2011}
\begin{equation}
\hat{a}_{j}^{\dagger}=\hat{f}_j^{\dagger}\prod_{m=1}^{j-1}e^{-i\pi \hat{f}_m^{\dagger}\hat{f}_m},~ \hat{a}_{j}=\prod_{m=1}^{j-1}e^{i\pi \hat{f}_m^{\dagger}\hat{f}_m}\hat{f}_j,
\end{equation}
where $\hat{f}_j^{\dagger}$ and $\hat{f}_j$ are the creation and annihilation operators for free fermions, respectively. After this mapping, the pump induced by the adiabatic cyclic modulation $\beta$ are related to a 2D parameter space spanned by the parameters $t_1$, $t_2$, $\mu_A$ and $\mu_B$ with the following relations:
\begin{equation}
t_1-t_2=-\cos{\beta},~\frac{\mu_A-\mu_B}{2}=g+\sin{\beta}\sin{\phi}.
\end{equation}
Here we have taken $t=1$, $h=1$ and $\delta=-0.5$ throughout the pump cycle. Since the quantized pump depends only on the topology of the pump path, as shown in Fig. \ref{fig:5}(b), we consider three typical pump paths corresponding to $C_{\text{MB}}=0,\pm1$ in the topological phase diagram in Fig. \ref{fig:2}. The nontrivial or trivial pump depends on whether the trajectory encloses the degenerate point in the parameter plane or not. The pump cycle which does not enclose the origin point at $t_1-t_2=\frac{\mu_A-\mu_B}{2}=0$ corresponds to $C_{\text{MB}}=0$. On the contrary, the sequences with winding paths result in topological pumping with $C_{\text{MB}}=\pm1$, where the sign is determined by the pump direction of the particles.

For soft bosons in the topological Mott insulator phase, we numerically find that the topological pumping under OBCs remains for $U\gtrsim8$. In this regime, we can obtain a one-unit-cell jump in $X_{\text{COM}}(\beta)$, such as the results for $U=8$ shown in Figs. \ref{fig:6}(a) and \ref{fig:6}(c), which corresponds to the shift of an edge mode from one side to the other side. However, the topological pumping would be broken when the interaction is much smaller, which can be seen from $X_{\text{COM}}(\beta)$ for $U=3$ in Figs. \ref{fig:6}(a) and \ref{fig:6}(c). The breakdown of the topological pumping of edge modes for weakly interacting bosons can be understood from the corresponding energy spectra under OBCs, as shown in Figs. \ref{fig:6}(b) and \ref{fig:6}(d). In this case, the two middle branches of quasiparticles are no longer edge modes and do not connect the lower and the upper bulk spectra when $\beta$ continuously varies from $-\pi$ to $\pi$. In our numerical simulations, the breakdown of topological pumping when decreasing the interaction strength from $U=8$ to $U=3$ is a crossover instead of a phase transition, with $\delta X_{\text{COM}}$ being not quantized (and nearly zero). This is consistent with numerical results of the energy spectra, where the edge modes gradually mix into the bulk when decreasing $U$. In the case of $U\approx3$, one may detect the Chern number from the interferometric measurements to reveal the topological nature of the system \cite{Atala2013,DWZhang2018}. The topological properties of soft bosons in the weakly interacting regime and the related experimental measurements require future studies.

\subsection{Topological phase transition}

\begin{figure}[htbp]
\centering
\includegraphics[width=8.5cm]{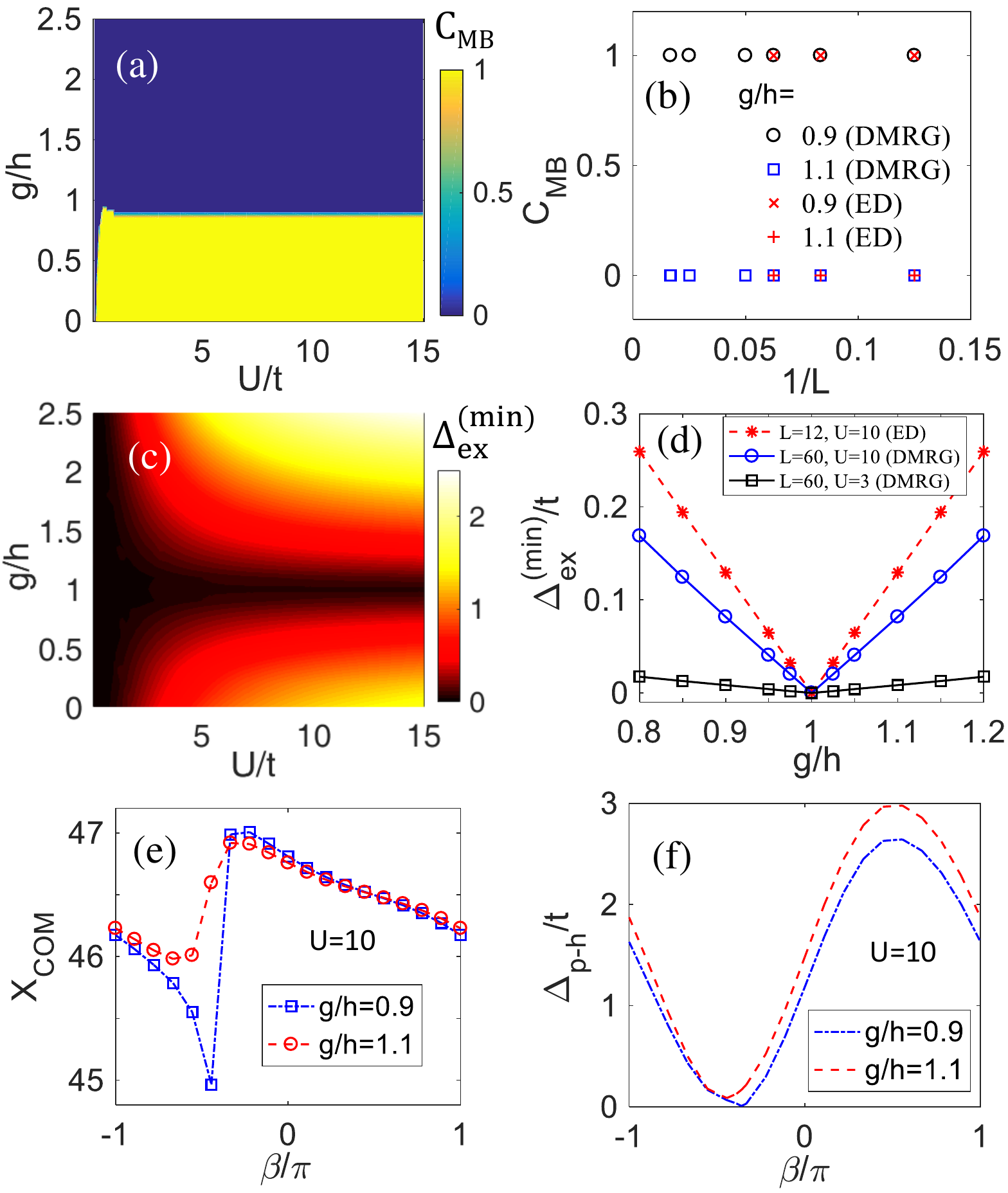}
\caption{(Color online). (a) The many-body Chern number $C_{\text{MB}}$ as a function of $U$ and $g/h$ for $L=12$ from ED. (b) $C_{\text{MB}}$ as a function of $1/L$ for $U=10$ and $g/h=0.9,1.1$ from ED and DMRG. (c) The minimum excitation gap $\Delta_{\text{ex}}^{\text{(min)}}$ (under the twisted PBC) as a function of $U$ and $g/h$ for $L=12$ from ED. (d) $\Delta_{\text{ex}}^{\text{(min)}}$ as a function of $g/h$ for $U=10$ with $L=12,60$ and $U=3$ with $L=60$. (e) The center-of-mass $X_{\text{COM}}$ in two pumps for $g/h=0.9,1.1$. (f) The particle-hole gap $\Delta_{p-h}$ (under OBCs) as a function of $\beta$ for $U=10$ and $g/h=0.9,1.1$. The other parameters are $h=1$,  $\delta=-0.5$, and $\phi=-\pi/2$.}
\label{fig:7}
\end{figure}

We now proceed to study the topological phase transition between normal ($C_{\text{MB}}=0$) and topological ($C_{\text{MB}}=1$) Mott insulators. To do this, we first calculate the many-body Chern number $C_{\text{MB}}$ as a function of the interaction strength $U$ and the system parameter $g/h$ (with fixed $\phi=-\pi/2$ and $\delta=-0.5$) for the lattice with length $L=12$, as shown in Fig. \ref{fig:7}(a). We find that in the Mott insulator regime, $C_{\text{MB}}=1$ is independent of $U$ as long as $g/h<1$, while $C_{\text{MB}}=0$ when $g/h>1$. This result can be understood from the hardcore bosons at half filling, which can be mapped to free fermions acting as a topological (trivial) band insulator with nonzero (zero) Chern numbers when $g/h<1$ ($g/h>1$). With decreasing interactions (as long as $U>U_c$), due to the gapped nature of the Mott insulators, the corresponding topological Chern numbers protected by the many-body gap still remain unchanged and then are independent on the interactions \cite{SLZhu2013,XDeng2014,Kuno2017}. Thus, the topological phase diagrams obtained from the Chern numbers preserve for $U=10$ and $U=3$, as shown in Fig. \ref{fig:2}. We also check that $C_{\text{MB}}$ is independent of the system-size $L$ by calculating $C_{\text{MB}}$ from $L=8$ to $L=60$ using both ED and DMRG, with two examples shown in Fig. \ref{fig:7}(b).

\begin{figure}[htbp]
\centering
\includegraphics[width=6cm]{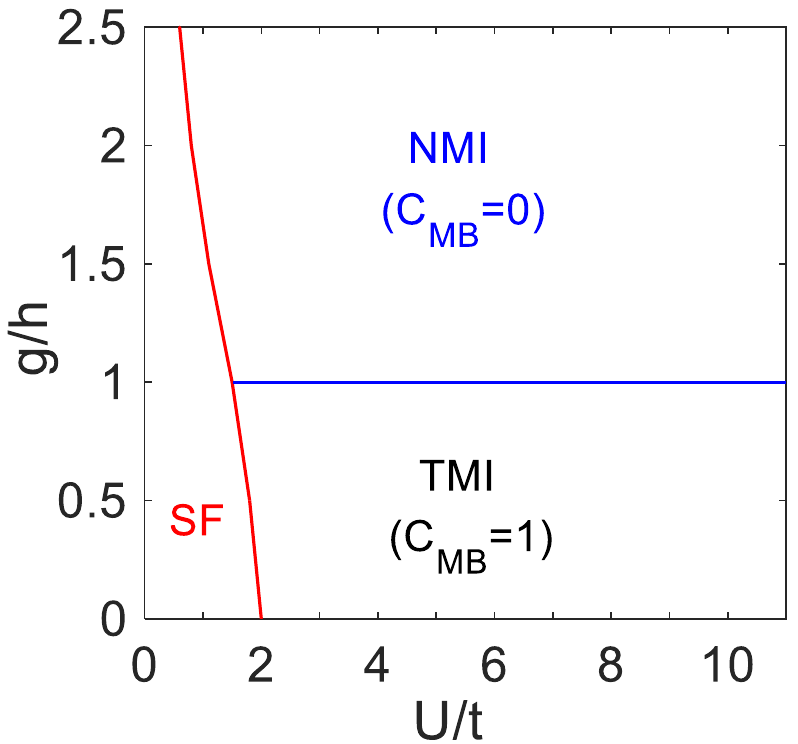}
\caption{(Color online). The global phase diagram for $U$ and $g/h$. Here SF, NMI and TMI denote the superfluid phase, the normal Mott insulating phase characterized by $C_{\text{MB}}=1$, and the topological Mott insulating phase characterized by $C_{\text{MB}}=1$, respectively. Other parameters are $h=1$, $\delta=-0.5$, and $\phi=-\pi/2$.}
\label{fig:8}
\end{figure}

The topological phase transition can be linked to the closing of the bulk excitation gap $\Delta_{\text{ex}}$ in the $\theta$-$\beta$ parameter space. To reveal the gap-closing nature of topological transitions, we calculate the minimums of $\Delta_{\text{ex}}$ in the parameter space under the twisted PBC:
\begin{equation}
\Delta_{\text{ex}}^{\text{(min)}}=\min[\Delta_{\text{ex}}(\theta,\beta)].
\end{equation}
As shown in Fig. \ref{fig:7}(c), the gap $\Delta_{\text{ex}}^{\text{(min)}}$ apparently closes along the line of $g/h=1$ for all $U>U_c$, which corresponds the change of the many-body Chern numbers of the ground state shown in Fig. \ref{fig:7}(a). In Fig. \ref{fig:7}(d), to show the gap-closing more clearly, we plot $\Delta_{\text{ex}}^{\text{(min)}}$ as a function of $g/h$ for $U=10$ with $L=12,60$ and $U=3$ with $L=60$, respectively. Here $\Delta_{\text{ex}}^{\text{(min)}}$ scales linearly with $g/h$ away from the transition at $g/h=1$, which indicates that the lattice potential offset $g/h$ (here $h=1$ is set) contributes to a (trivial) single-particle gap for interacting bosons in the Mott insulating phase. We also calculate the center-of-mass $X_{\text{COM}}$ in two typical pumps for $g/h=0.9$ and $1.1$, respectively, as shown in Fig. \ref{fig:7}(e). For the topological (trivial) pumping when $g/h=0.9$ ($g/h=1.1$), the results show one (zero) unit-cell jump in $X_{\text{COM}}(\beta)$, which indicate the shift of $|C_{\text{MB}}|=1$ ($|C_{\text{MB}}|=0$) edge mode from one side of the lattice to the other under OBCs. As shown in Figs. \ref{fig:4}(c) and \ref{fig:4}(d), the two branches of edge modes in the topological phase cross and connect the bulk spectrum, as one continuously varies the parameter $\beta$.
Under OBCs (note that $\theta$ is just defined under the twisted PBC), we can define the particle-hole gap of the quasiparticles as \cite{XDeng2014}
\begin{equation}
\Delta_{p-h}=\delta E^{(O)}_{L/2}-\delta E^{(O)}_{L/2-1}
\end{equation}
near the half filling, where the quasiparticle energy $\delta E^{(O)}_{n}$ is given in Eq. (\ref{En}). As shown in Fig \ref{fig:7}(f), $\Delta_{p-h}(\beta)$ closes for $g/h=0.9$ with $C_{\text{MB}}=1$ at a $\beta$ value, while $\Delta_{p-h}(\beta)$ is open for $g/h=1.1$ with $C_{\text{MB}}=0$.

Previously, we have studied the topological properties of the ground state of the system and revealed the topological phase transition between normal and topological Mott insulators when $g/h=0$ for $U>U_c$. In addition, for $g/h=0$ and $0.5$, the ground state undergoes a quantum phase transition from the superfluid to the Mott insulators at the critical interaction strengths $U_{c}\approx2$ and $1.8$, respectively, which are obtained by the finite-size scaling analysis of the excitation gap. Using the same method, we can determine $U_{c}$ for other values of $g/h$, and then obtain the global phase diagram with respect to $g/h$ and $U$, as shown in Fig. \ref{fig:8}. The phase diagram shows the critical $U$ and $g/h$ for determining the boundaries of three quantum phases: superfluid, normal Mott insulator with $C_{\text{MB}}=0$, and topological Mott insulator with $C_{\text{MB}}=1$. One can find that the value of the critical point $U_{c}$ is decreased with the increasing of $g/h$ since the potential offset increases the band gap in the single-particle energy bands.

\section{conclusion}\label{sec4}


In summary, we have studied an extended 1D Bose-Hubbard model at half filling and shown that the ground state in the strong interaction regime can simulate bosonic Chern insulators with a topological phase diagram similar to that of the 2D Haldane model. Moreover, we have explored the topological properties by calculating the many-body Chern numbers, the in-gap edge states in the quasiparticle energy spectrum, the topological pumping, and the topological phase transition between normal (trivial) and topological Mott insulators. We also have obtained the global phase diagram with the superfluid and the two Mott insulating phases. Our proposed model could be realized with ultracold bosonic atoms in a 1D dimerized optical superlattice with tunable parameters and atomic interactions. The topological properties in the system, such as the topological pumping, can be detected in current cold-atom experiments, allowing for further exploration of interacting topological phases with ultracold atoms.

\begin{acknowledgments}
We thank Liang He for helpful discussions. This work was supported by the NKRDP of China (Grant No. 2016YFA0301800), the NSFC (Grants No. 11604103), the NSAF (Grant No. U1830111), the Key-Area Research and Development Program of Guangdong Province (Grant No. 2019B030330001), and the Key Project of Science and Technology of Guangzhou (Grant No. 201804020055).
\end{acknowledgments}

\bibliography{reference}

\end{document}